\documentclass[12pt,a4paper]{article}
\usepackage[T2A]{fontenc}
\usepackage[utf8]{inputenc}
\usepackage[english]{babel}
\usepackage{amssymb}
\usepackage{amsmath}
\usepackage{graphicx}
\usepackage[small,sc,center]{titlesec}
\usepackage{indentfirst}
\usepackage{siunitx}
\usepackage[labelsep=period]{caption}
\usepackage{caption}

\begin{document}

\begin{center}
	\textbf{Models of Mira variables of the Large Magellanic Cloud}

	\vskip 3mm
	\textbf{Yu. A. Fadeyev\footnote{E--mail: fadeyev@inasan.ru}}

	\textit{Institute of Astronomy, Russian Academy of Sciences,
    		Pyatnitskaya ul. 48, Moscow, 119017 Russia} \\

	Received September 23, 2024; revised October 24, 2024; accepted October 24, 2024
\end{center}

\textbf{Abstract} ---
Consistent stellar evolution and nonlinear radial stellar pulsation calculations were carried out for
models of asymptotic giant branch stars with initial masses $1.5M_\odot\le M_\mathrm{ZAMS}\le 3M_\odot$
and initial metal abundance $Z=0.006$.
All the models are shown to be either the fundamental mode or the first overtone pulsators.
The lower limit of the first overtone period increases with increasing mass of the Mira model from
$\Pi_{1,\min}\approx 80$~days for $M=1.3M_\odot$ to $\Pi_{1,\min}\approx 120$~days for $M=2.6M_\odot$.
The upper limit of the first overtone period and lower limit of the fundamental mode period depend
on the stellar structure during mode switching and range from $\Pi_{1,\max}=130$,
$\Pi_{0,\min}=190$~days for $M=0.96M_\odot$ to $\Pi_{1,\max}=210$, $\Pi_{0,\min}=430$~days for
$M=2.2M_\odot$.
The slope of the theoretical period--luminosity relation of Mira variables perceptibly increases with
decreasing $Z$.
Fourier spectra of the kinetic energy of twelve hydrodynamic models show a split of the fundamental
mode maximum into several equidistant components.
Frequency intervals between split components fall within the range $0.03 \le \Delta\nu/\nu_0 \le 0.1$.
The superposition of radial oscillations with the fundamental mode splitting leads to the long--term
amplitude variations with the cycle length from 10 to 30 times longer than the fundamental mode period.
A more thorough analysis of hydrodynamic models is required for understanding the origin of the
principal pulsation mode splitting.


\newpage
\section*{introduction}

The period--luminosity relation of Mira variables discovered by Glass and Lloyd~Evans (1981) from IR
observations of red giants in the Large Magellanic Cloud (LMC) is now a reliable tool for determination
of interstellar distance scale (Glass and Feast 1982; Feast 1984; Whitelock et al. 2000; 2008;
Iwanek et al. 2021).
Moreover, observations of long period variables in neighbouring galaxies allow us to expand limits of
the usage of Mira variables as distance indicators to the Local Group of galaxies
(Mould 2004; Whitelock et al. 2013; Huang et al. 2018; Yan et al. 2018).
At the same time, despite of significant progress in observations of Mira variables
their physical nature is still poorly understood.
First of all, one should note uncertainties in the theory of late stages of stellar evolution because the
Mira variables are the stars of the asymptotic giant branch undergoing thermal pulses (i.e. TP--AGB stars)
in the helium shell source (Iben and Renzini 1983; Herwig 2005).
Another obstacle is that the application of the linear theory of stellar pulsation may lead to
dubious results because of the large oscillation amplitude in Mira variables.
Strong shock waves appearing because of the large radial displacement of outer layers are responsible
for significant redistribution of gas density in the stellar atmosphere and favour dust grain formation
(Willson 2000).
Models of red giant nonlinear pulsations computed with the numerical methods of radiation hydrodynamics
encounter significant difficulties when the expansion velocity of outer layers exceeds the local escape
velocity (Wood 1974; Tuchman et al. 1978; Olivier and Wood 2005).

Among the few studies devoted to nonlinear oscillations of red giants we have to note the paper by
Trabucchi et al. (2021) where a great deal of hydrodynamic models with various stellar masses, radii
and luminosities are successfully compared with Mira variables of the LMC.
Unfortunately, a significant shortcoming of this study is due to the fact that the initial conditions
required for solution of the equations of hydrodynamics were calculated with the carbon core mass -- luminosity
relation (Trabucchi et al. 2019) rather than from the stellar evolution computations.
In particular, this method does not allow to obtain the theoretical period--luminosity relation ($\Pi-L$).

Theoretical estimates of various characteristics of Mira variables as well as the $\Pi-L$ relation can be
obtained only on the basis of consistent computations of stellar evolution and nonlinear stellar pulsations
when initial conditions required for solution of the equations of hydrodynamics describing stellar
oscillations are determined using computed in advance evolutionary sequence models.
This method was earlier applied by the author (Fadeyev 2023) to determine the theoretical period--radius
and period--luminosity relations of Mira models with solar metallicity ($Z=0.014$).

Below we present the results of consistent computations of stellar evolution and nonlinear radial stellar
pulsations of Mira models with initial metal abundance $Z=0.006$ corresponding to the recent
observational estimates of metallicity of LMC stars (Rolleston et al. 2002; Cole et al. 2005).
The goal of this study implies the solution of the following problems:
\begin{itemize}
 \item Determination of the stellar evolution phases corresponding to the certain radial oscillation mode.
 \item Estimation of period ranges for each of oscillation modes.
 \item Determination of the theoretical $\Pi-L$ relation for Mira models with the metal abundance
       $Z=0.006$ and comparison with results obtained earlier for $Z=0.014$ (Fadeyev 2023).
\end{itemize}

Looking ahead, it should be noted that during this work we found that limit cycle oscillations of
some hydrodynamic models are characterized by frequency split of the principal oscillation mode
and show cyclic amplitude variations on the time scale by an order of magnitude longer than the
pulsation period.
In the final part of this paper we discuss both main features of this phenomenon and possible
explanation of the long secondary periodicity (LSP) discovered more than twenty years ago in pulsating
red giants of LMC (Wood et al. 1999; Wood 2000) but has not been explained yet.

\section*{evolutionary sequences of agb stars}

Stellar evolution from the main sequence up to the final AGB stage was computed with the program MESA
version r15140 (Paxton et al. 2019).
Details of computations and required parameters are discussed in our previous paper (Fadeyev 2023).
Altogether, we computed 9 evolutionary sequences with zero age main sequence masses
$1.5M_\odot\le M_\mathrm{ZAMS}\le 3M_\odot$ for initial abundances of helium and heavier elements
(metals) $Y=0.28$ and $Z=0.006$.

\begin{figure*}[t!]
\centering
\includegraphics[width=0.8\textwidth]{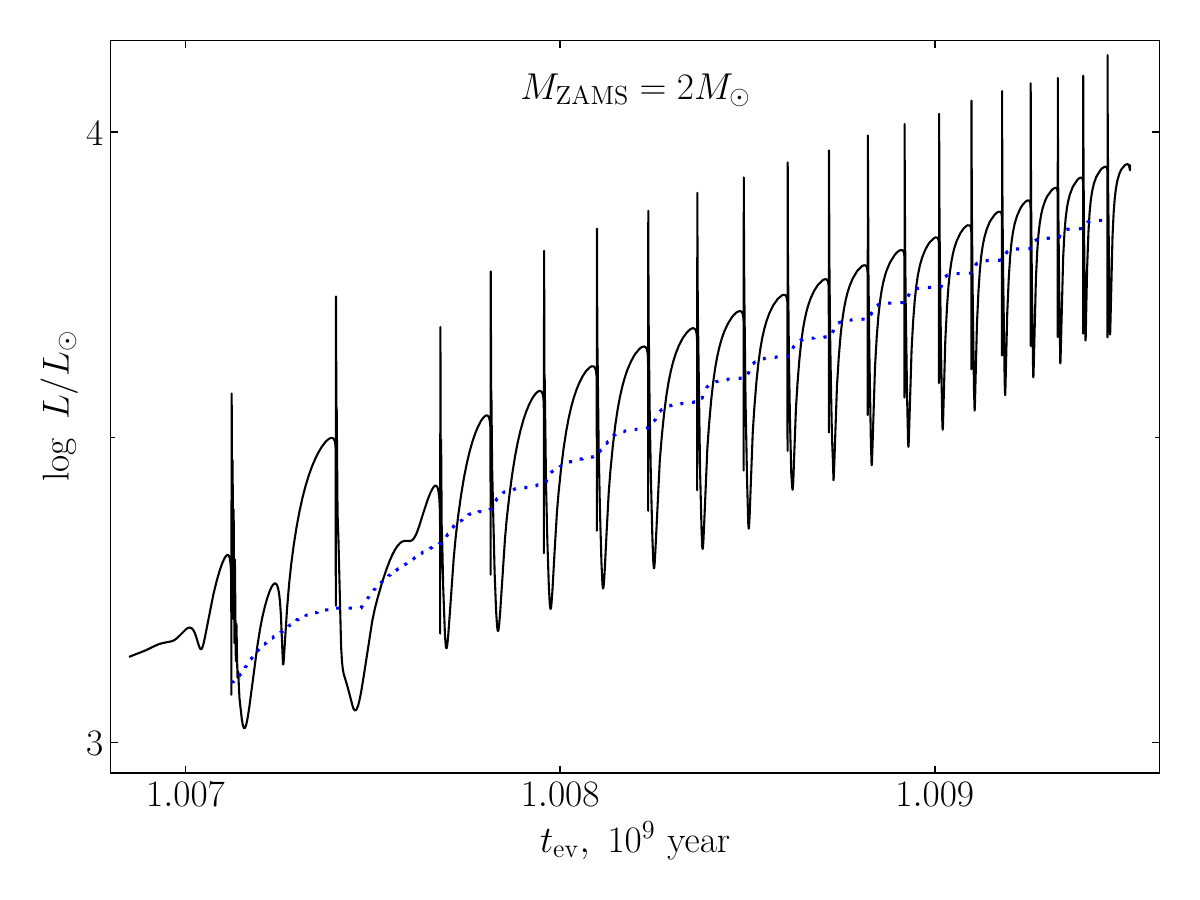}
\caption{Luminosity variations during the TP--AGB evolutionary stage in the star with initial mass
         $M_\mathrm{ZAMS}=2M_\odot$.
         The carbon core mass -- luminosity relation (\ref{mc-l}) is shown by the dotted line.\hfill}
\label{fig1}
\end{figure*}

The main feature of the TP--AGB evolutionary stage is that the luminosity of the helium shell source
experiences quasi--periodic increases accompanied by significant variations of the stellar radius and
luminosity that are responsible for secular changes of the pulsation period in some Mira variables
(Wood and Zarro 1981).
The typical plot of luminosity variations during $\approx 2.3\times 10^6$ yr of the TP--AGB phase
in the red giant with initial mass $M_\mathrm{ZAMS}=2M_\odot$ is shown in Fig.~\ref{fig1}.

The carbon core mass -- luminosity relation for the TP--AGB evolutionary sequence $M_\mathrm{ZAMS}=2M_\odot$
determined by the weighted least squares with weights proportional to the evolution time can be approximately
written as
\begin{gather}
\label{mc-l}
L/L_\odot = 4.255\times 10^4 \left(M_\mathrm{C} - 0.426\right)
\end{gather}
and is shown in Fig.~\ref{fig1} by the dotted line which deviates from the straight line due to
non--monotonic growth of the carbon core.
As seen in Fig.~\ref{fig1}, each thermal flash is accompanied by significant variations of the
stellar luminosity so that the use of (\ref{mc-l}) for calculation of initial conditions will inevitably
lead to wrong results.

In the present study the initial conditions required for the solution of the equations of
hydrodynamics describing stellar pulsations were calculated using the selected TP--AGB stellar
models.
One should bear in mind that not all models of the TP--AGB evolutionary sequence can be
used for calculation of initial conditions since the applicability of the theory of stellar pulsation
is restricted to the conditions of hydrostatic and thermal equilibrium.
The first of these conditions is fulfilled in all evolutionary models because it is a part
of the solution of stellar structure equations.
At the same time during some evolutionary phases the thermal imbalance can appear in the stellar
envelope when the gravitational energy changes because of contraction or expansion of the star.
In particular, such changes take place in TP--AGB red giants during short--term maxima of the
helium shell source luminosity.
The duration of the thermal imbalance in Mira variables does not exceed a few percent of the
interflash interval (Fadeyev 2023) and on the luminosity plot in Fig.~\ref{fig1} this time
interval is in the close vicinity of luminosity peaks.
In the present study the calculations of initial conditions were restricted to the models
of evolutionary sequences corresponding to the monotonic luminosity growth (see Fig.~\ref{fig1})
when the condition of thermal equilibrium is fulfilled to high accuracy.

It should be noted that the consistency between the initial conditions and calculations of
stellar pulsation is provided not only by distributions of the radius, luminosity and other
variables with respect to the Lagrangian mass coordinate but also by the same elemental abundances
within the stellar envelope.
This condition is of importance during the TP--AGB stage because of changes of elemental
abundances after the onset of the 3rd dredge--up from the helium--burning shell.

\section*{hydrodynamic models of miras}

The basic equations of radiation hydrodynamics and time--dependent convection we used for
calculation of radial oscillations in Miras are discussed in one of our preceding papers
(Fadeyev 2013).
There are two types of the Cauchy problem solution for nonlinear stellar pulsations.
The first describes the amplitude growth and subsequent transition to oscillations
with the constant amplitude.
The second describes decaying oscillations indicating that the stellar model is
stable against radial pulsations.
For both types of the solution we calculated the pulsation period $\Pi$ using the discrete
Fourier transform of the kinetic energy of stellar pulsation motions.
In the case of pulsationally unstable models the period was calculated after attainment
of  limiting amplitude whereas in the case of decaying oscillations the period was
determined for the whole interval of the solution.
The relative error of the period determination does not exceed $\approx 1\%$.

\begin{figure*}[t!]
 \centering
\includegraphics[width=0.8\textwidth]{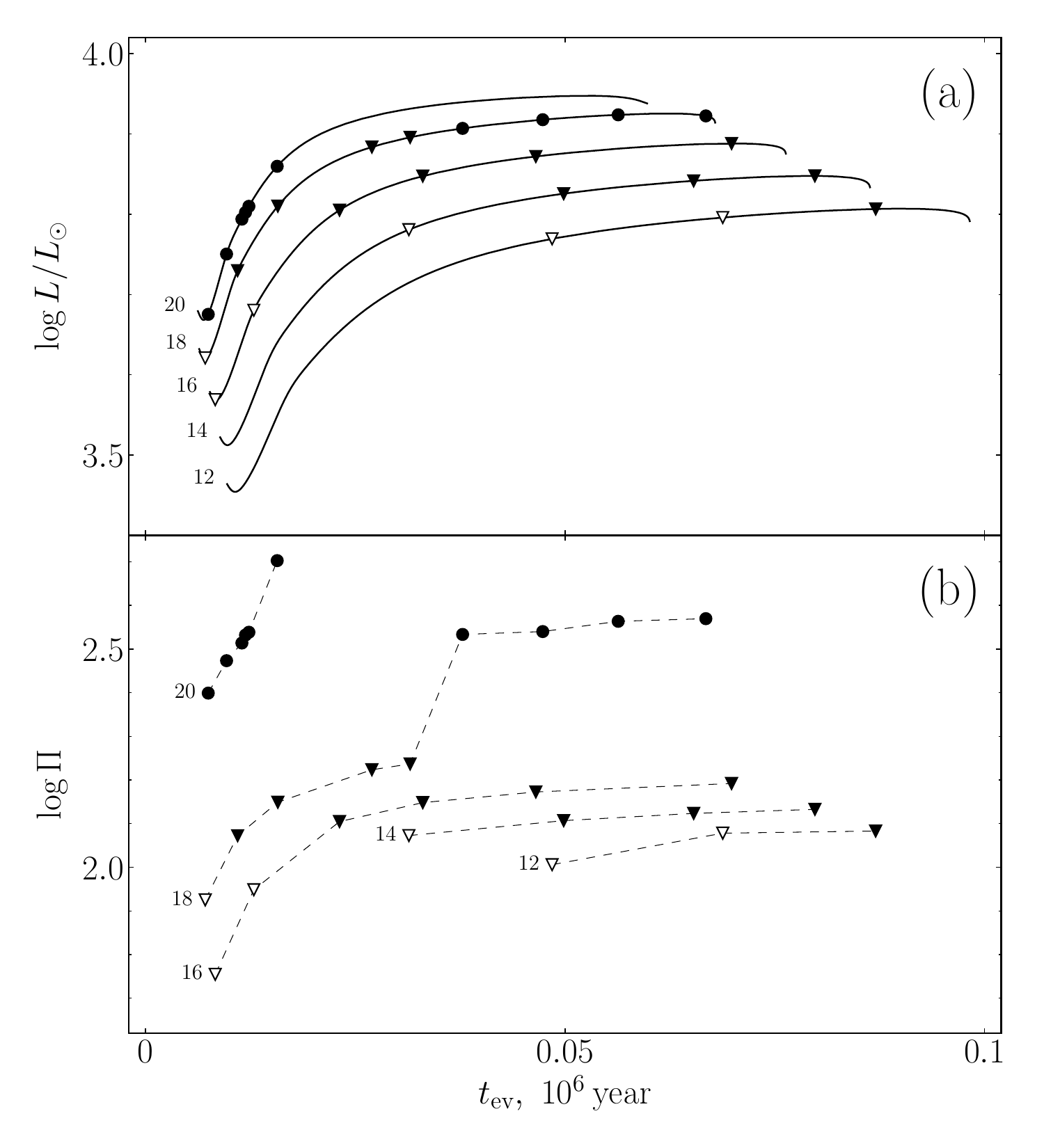}
\caption{Variations of the luminosity (a) and period of radial oscillations (b) after the
         helium flash in models of the evolutionary sequence $M_\mathrm{ZAMS}=2M_\odot$ during
         luminosity growth.
         On the left of the plots the thermal flash number $i_\mathrm{TP}$ is marked.
         The evolution time is set to zero at the luminosity peak of the helium shell source.
         Hydrodynamic models pulsating in the fundamental mode and first overtone are shown
         by circles and triangles.
         The filled and open symbols correspond to models of pulsating stars and to those
         with decaying oscillations.}
\label{fig2}
\end{figure*}

Results of consistent stellar evolution and nonlinear stellar pulsation calculations for the
evolutionary sequence $M_\mathrm{ZAMS}=2M_\odot$ are illustrated in Fig.~\ref{fig2}.
For the sake of convenience the results are plotted for thermal flashes with even numbers
$12\le i_\mathrm{TP}\le 20$ and the evolution time $t_\mathrm{ev}$ is set to zero at the
maximum peak of the helium–-shell luminosity
Circles and triangles mark hydrodynamic models pulsating in the fundamental mode and first
overtone whereas open symbols correspond to models with decaying oscillations.

As seen in Fig.~\ref{fig2}, during the initial TP--AGB stage (i.e. for $i_\mathrm{TP} < 12$) the
models of the evolutionary sequence $M_\mathrm{ZAMS}=2M_\odot$ are stable against the radial
oscillations.
However the hydrogen and helium ionization zones, where pulsations are driven, expand as the luminosity
increases so that the star begins to pulsate after the thermal pulse $i_\mathrm{TP} = 12$.

At the beginning of oscillations the star is the first overtone pulsator because the inner boundary
of the hydrogen and helium ionization zones locate above the node of the first overtone.
The radius of the node is $r_\mathrm{n}\approx 0.77R$, where $R$ is the surface radius of the evolution
model.
The duration of the evolutionary stage when the star pulsates increases with increasing number
of the thermal flash because of higher stellar luminosity and the larger size of ionization
zones.
As seen in Fig.~\ref{fig2}a, for $12\le i_\mathrm{TP}\le 16$ the star pulsates in the first overtone.
For $i_\mathrm{TP}=18$ the growth of stellar luminosity is accompanied by pulsation mode switch
from the first overtone to the fundamental mode since the inner boundary of ionization zones plunges
below the first overtone node.
The thermal flash $i_\mathrm{TP} = 20$ is the last in this evolutionary sequence.
The low stellar mass ($M\le 1.26 M_\odot$) and increasing luminosity are responsible for
further increase of the pulsation amplitude and less regular oscillations.

As seen in Fig.~\ref{fig2}b, during the final TP--AGB stage of the stellar evolution
the range of periods variation during the thermal pulse cycle significantly increases.
For example, during the cycle $i_\mathrm{TP} = 18$ the pulsation periods changes from
$\approx 120$~days (oscillations in the first overtone) to $\approx 610$~days (fundamental mode
pulsations).

The pulsation period monotonically increases during each cycle of thermal instability so that
we can evaluate the minimum and maximum values of the first overtone and fundamental mode periods
for each cycle of thermal instability and ultimately to determine the limits of period changes for
the evolutionary sequence with initial mass $M_\mathrm{ZAMS}$.
In particular, the first overtone period at the edge of pulsational instability is determined as
the mean value of the periods of two adjacent models one of which shows decaying oscillations whereas
another is unstable against radial pulsations.
The upper limit of the first overtone period and the lower limit of the fundamental mode period
are evaluated in vicinity of the mode switch (see, for example, the plot $i_\mathrm{TP}=18$ in
Fig.~ \ref{fig2}b):
\begin{gather}
\label{p1max}
\Pi_{1,\max} = \frac{1}{2} (\Pi_1 + \frac{1}{2}\Pi_0) ,
\end{gather}
\begin{gather}
\label{p0min}
\Pi_{0,\min} = \frac{1}{2} (2\Pi_1 + \Pi_0) ,
\end{gather}
where $\Pi_1$ and $\Pi_0$ are the first overtone and fundamental mode periods of two adjacent models.
In relations (\ref{p1max}) and (\ref{p0min}) we used the fact that the periods of the first overtone and
fundamental mode relate as $\Pi_1/\Pi_0 = 1/2$.

\begin{figure*}[t!]
 \centering
\includegraphics[width=0.8\textwidth]{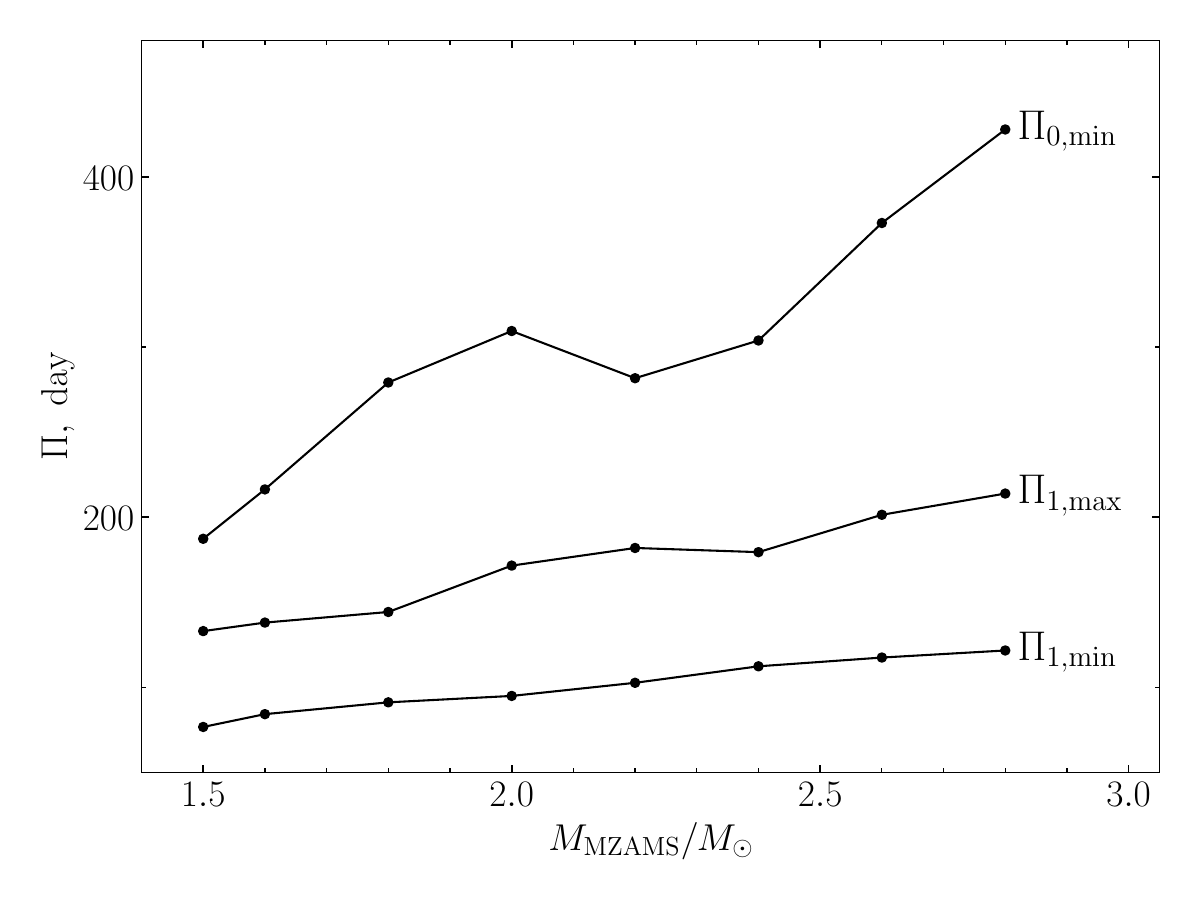}
\caption{The minimum and maximum values of the first overtone period $\Pi_{1,\min}$ and
        $\Pi_{1,\max}$ and the minimum value of the fundamental mode period $\Pi_{0,\min}$
        against the initial stellar mass $M_\mathrm{ZAMS}$.}
\label{fig3}
\end{figure*}

Rough estimates of the limit values of pulsation periods in Mira models are shown in Fig.~\ref{fig3}
as a function of the initial mass of the evolutionary sequence $M_\mathrm{ZAMS}$.
It should be noted that non--monotonic increase in $\Pi_{1,\max}$ and $\Pi_{0,\min}$ seems to be due to
insufficiently dense grid of hydrodynamic models for $2M_\odot\le M_\mathrm{ZAMS}\le 2.4M_\odot$.

\section*{period--luminosity relation}

\begin{figure}
 \centering
 \includegraphics[width=0.70\columnwidth,clip]{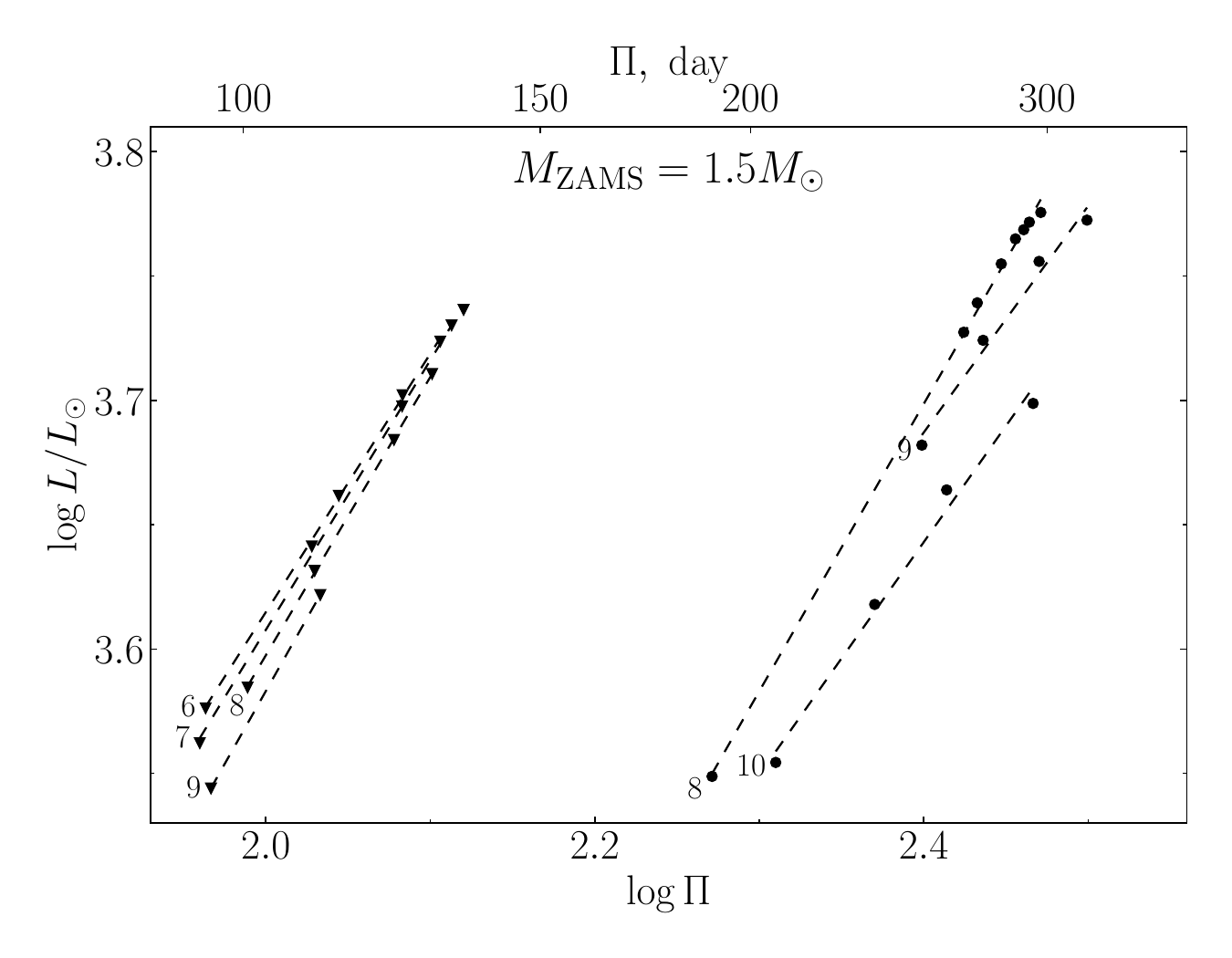}
 \caption{Hydrodynamic models of the evolutionary sequence $M_\mathrm{ZAMS}=1.5M_\odot$ on the
          diagram period -- luminosity. Circles and triangles represent the fundamental mode and
          first overtone pulsators.
          Dashed lines represent the least square fit for models of the same thermal cycle.
          Index of the thermal cycle $i_\mathrm{TP}$ is shown on the left at each dependence.}
\label{fig4}
\end{figure}

As shown above, during a cycle of thermal instability the pulsation period varies within a wide range
so that pulsations with the same period can appear at different values of $i_\mathrm{TP}$.
Therefore the luminosity of stars pulsating with nearly the same period can vary due to monotonic growth
of the carbon core mass as the star ascends the AGB and models of the same evolutionary sequence
are dispersed in the $\Pi-L$ diagram.
This feature is illustrated in Fig.~\ref{fig4}, where the period--luminosity diagram is shown for
hydrodynamic models of the evolutionary sequence $M_\mathrm{ZAMS}=1.5M_\odot$ pulsating in the
first overtone ($6\le i_\mathrm{TP}\le 9$) and in the fundamental mode ($8\le i_\mathrm{TP}\le 10$).
As can be seen from these plots, the values of $\log\Pi$ and $\log L$ agree with the linear fit
to good accuracy for a fixed value of $i_\mathrm{TP}$ but the fitting straight lines are shifted to
the right along the horizontal axis because of stellar mass decrease.
Stellar oscillations in the fundamental mode take place during stronger mass loss when deviation
of $\log\Pi$ and $\log L$ from the linear fit becomes more appreciable.

\begin{figure}
 \centering
 \includegraphics[width=0.60\columnwidth,clip]{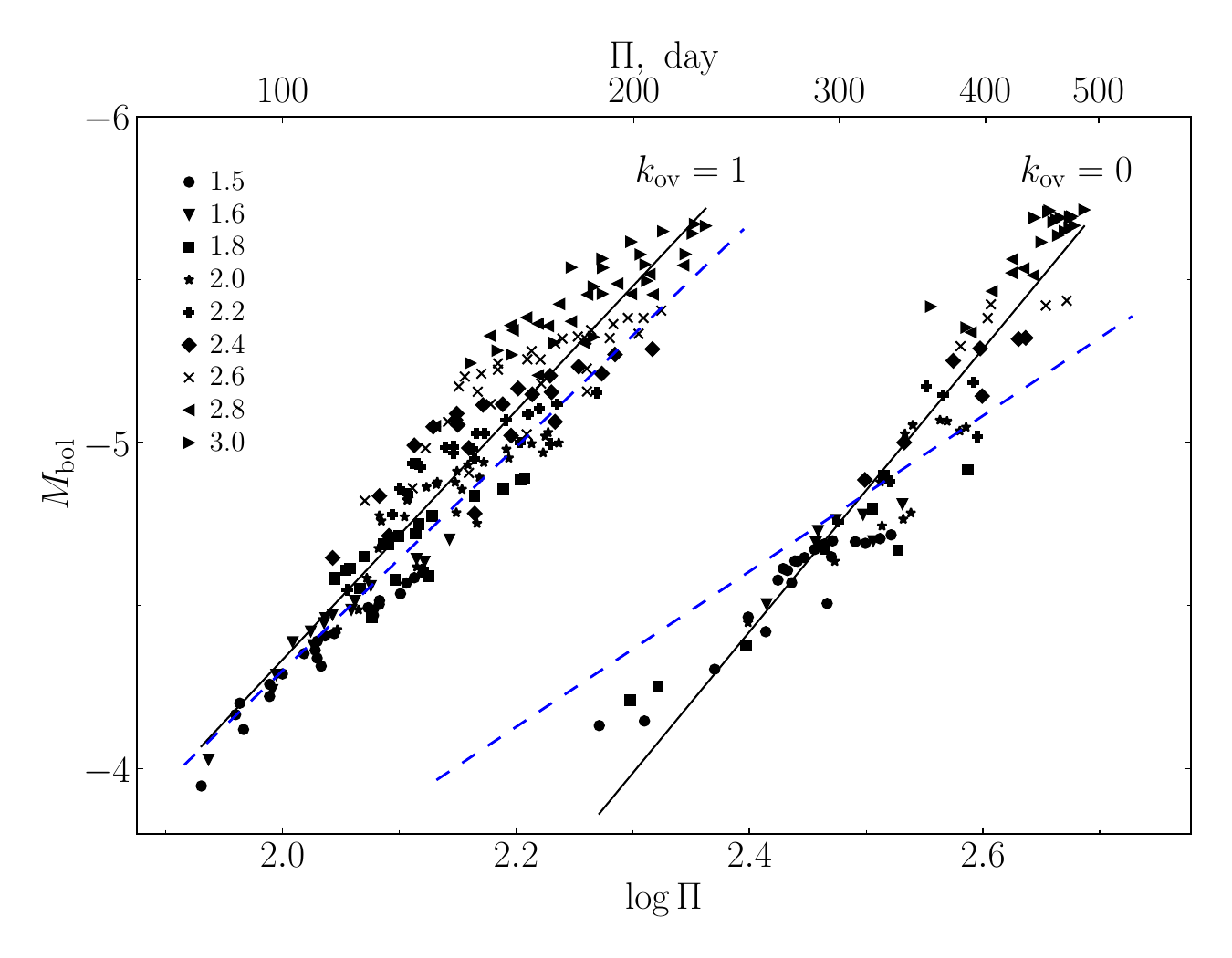}
 \caption{Period--luminosity relations for the fundamental mode ($k_\mathrm{ov}=0$)
          and first overtone ($k_\mathrm{ov}=1$) pulsators with metal abundance $Z=0.006$.
          Different symbols represent the models of evolutionary models with
          $1.5M_\odot\le M_\mathrm{ZAMS}\le 3M_\odot$.
          The solid lines represent linear fits (\ref{pl0}) and (\ref{pl1}), whereas the dashed lines
          show the linear fits obtained for Mira variable models with $Z=0.014$ (Fadeyev 2023).}
\label{fig5}
\end{figure}

In the present study we computed 195 hydrodynamic models of Mira variables pulsating in the first
overtone and 90 models pulsating in the fundamental mode.
The theoretical $\Pi-L$ relations for the fundamental mode and first overtone are shown in
Fig.~\ref{fig5}, where the different symbols correspond to models of evolutionary sequences
that are listed in the upper left corner of Fig.~\ref{fig5}.

The solid lines in Fig.~\ref{fig5} represent the linear fits of ($\log L$, $\lg\Pi_0$) and
($\log L$, $\lg\Pi_1$) values.
\begin{gather}
\label{pl0}
M_\mathrm{bol} = -4.328 \log\Pi_0 + 5.967 ,
\\
\label{pl1}
M_\mathrm{bol} = -3.820 \log\Pi_1 + 3.306 .
\end{gather}
A noticeable scatter of points around the straight lines (\ref{pl0}) and (\ref{pl1}) is due to
different values of the carbon core mass of hydrodynamic models.

The diagram in Fig.~\ref{fig5} as well as relations (\ref{pl0}) and (\ref{pl1}) correspond to the
absolute bolometric magnitude $M_\mathrm{bol}$ whereas all empirical period--luminosity relations
of Mira variables in the LMC were determined for near--infrared magnitudes.
For example, the slope of the $\Pi-L$ relation evaluated from observation in the K passband of
53 Mira variables in the LMC with periods from 116 to 413~days is $\rho = -3.69$ (Whitelock et al. 2008).
According to Josselin et al. (2000) the bolometric magnitudes of AGB stars and their K magnitudes,
corrected for interstellar extinction, relate as
\begin{gather}
\label{mbol-k}
m_\mathrm{bol} \simeq m_\mathrm{K} - 3 .
\end{gather}
There are two ways to explain the difference between the slope  $-4.33$ of theoretical relation (\ref{pl0})
and the value $-3.69$ of the empirical relation by Whitelock et al. (2008).
First, we have to assume that the bolometric correction relating $m_\mathrm{bol}$ and $m_\mathrm{K}$
is not constant.
Second, the better agreement between the theory and observations can probably be obtained for the larger
metal abundance $Z$.
In particular, the dashed lines in Fig.~\ref{fig5} that represent linear fits of the period--luminosity
relation for models with metal abundance $Z=0.014$ have smaller slope (-2.39 for fundamental mode
pulsators).
Therefore, the small increase of metal abundance in comparison with $Z=0.006$ can probably improve
agreement with observations.

\section*{models of mira variables with secondary periodicity}

\begin{figure}
 \centering
 \includegraphics[width=0.90\columnwidth,clip]{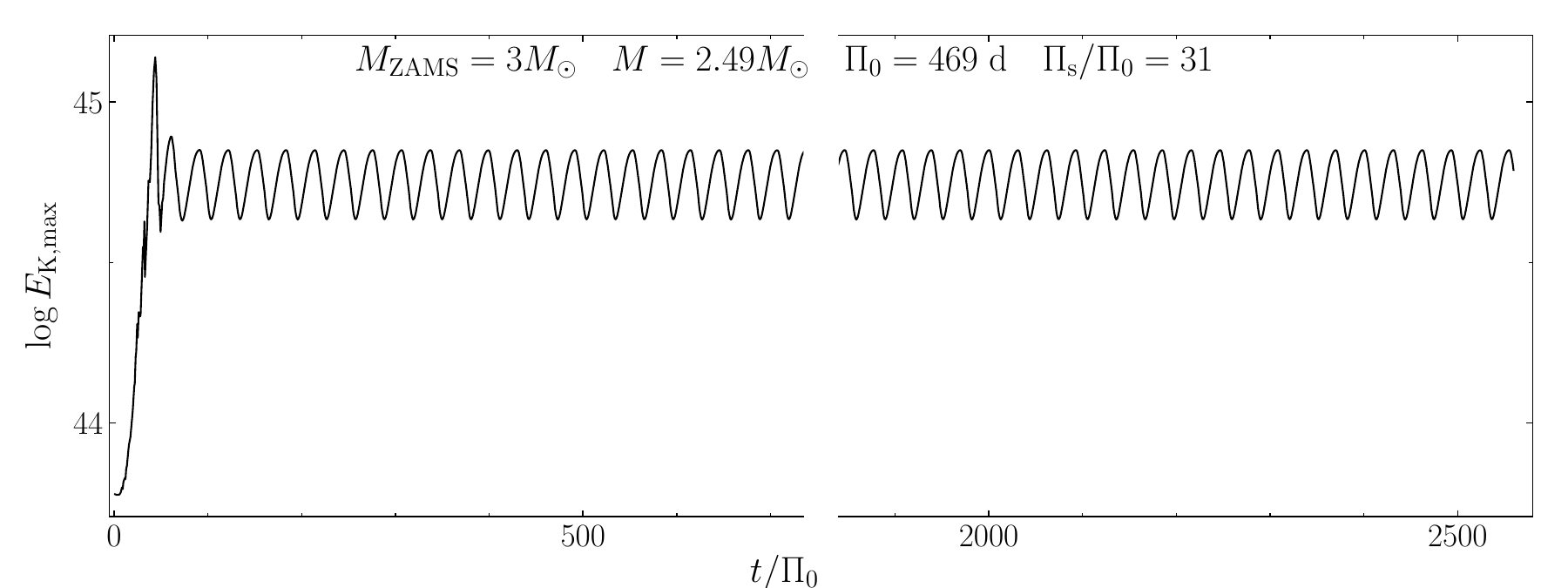}
 \caption{Variations of maximum values of the kinetic energy of pulsations motions in the
          model with mass $M=2.49M_\odot$ and the period $\Pi=469$~d
          (the evolutionary sequence $M_\mathrm{ZAMS}=3M_\odot$).}
\label{fig6}
\end{figure}

Among nearly three hundred models of Mira variables considered in the present study we found twelve
hydrodynamic models where the limit amplitude oscillations were accompanied by the long--term
secondary periodicity.
The example of radial pulsations with secondary periodicity is shown in Fig.~\ref{fig6}, where
variations of maximum values of the pulsation kinetic energy $E_{\mathrm{K},\max}$
are plotted against the number of fundamental mode periods $t/\Pi_0$ for the Mira variable model
with the mass $M=2.49M_\odot$ and period $\Pi_0=469$~days.
In models without the secondary periodicity the maximum of kinetic energy after attainment of the
limiting amplitude is independent of time $t$ whereas the model in Fig.~\ref{fig6} shows cyclic
variations of the pulsation amplitude on the time scale nearly 30 times longer than the period
of radial oscillations ($\Pi_\mathrm{s}/\Pi_0=31$).
Calculations of nonlinear pulsations for the model shown in Fig.~\ref{fig6} we carried out for
the long enough time interval ($t\sim 2.5\times 10^3\Pi_0$) but nevertheless attempts to detect
any secular changes of amplitude variations failed.

\begin{figure}
 \centering
 \includegraphics[width=0.60\columnwidth,clip]{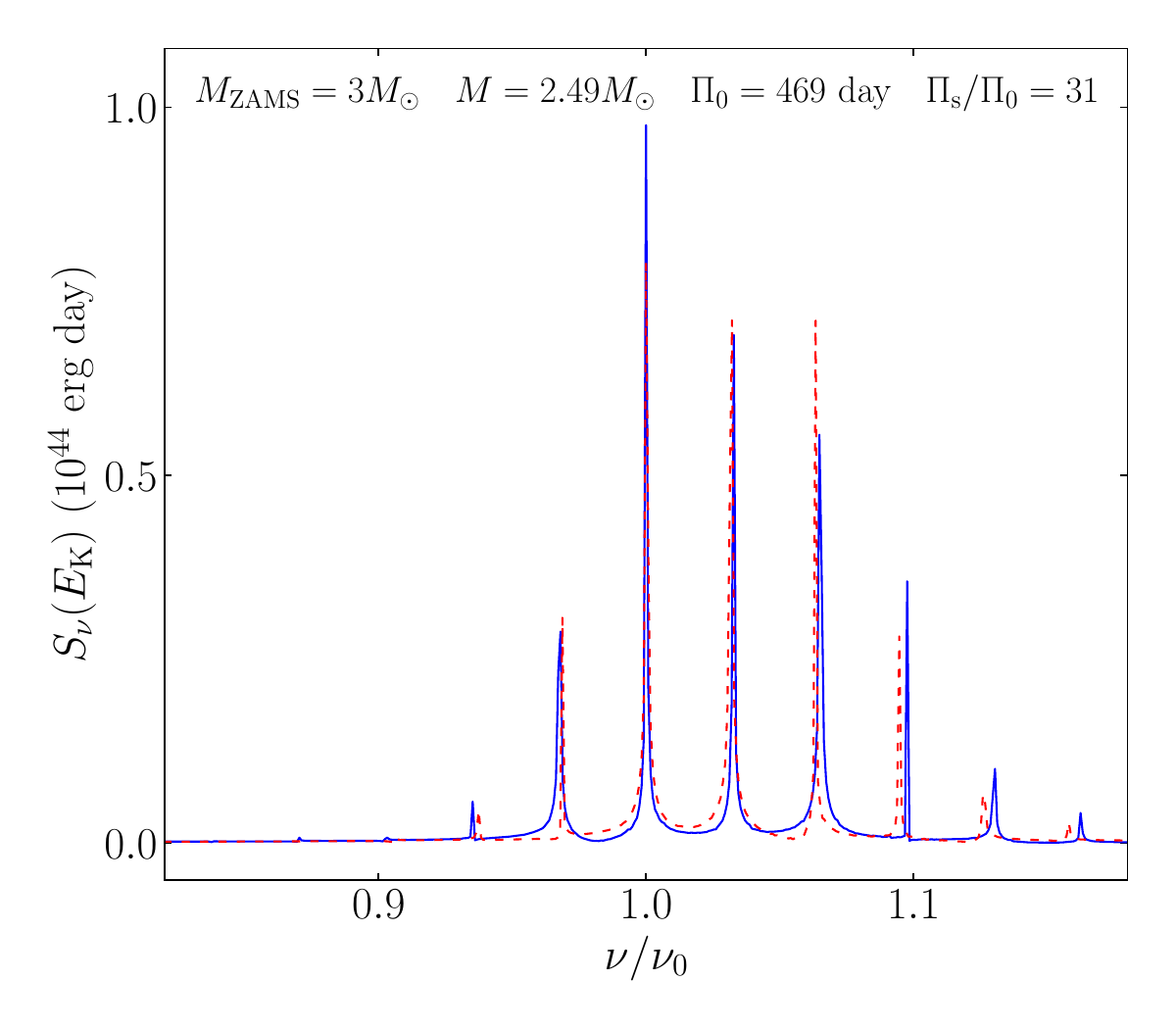}
 \caption{The power spectrum of the pulsation kinetic energy $S_\nu(E_\mathrm{K})$ in the vicinity
          of the fundamental mode frequency $\nu_0 = 1/\Pi_0$ of the hydrodynamic model shown in
          Fig.~\ref{fig6} (solid line).
          The power spectrum of the model with the doubled number of Lagrangean mass zones
          ($N=1200$) is shown by the dashed line.}
\label{fig7}
\end{figure}

To exclude any doubts that the secondary periodicity is an computational artefact arising due to
the discrete nature of the numerical model we carried out the nonlinear pulsation calculations
for the same model but with the doubled number of mass zones.
Results of calculations are shown in Fig.~\ref{fig7} where the kinetic energy power spectra are
plotted for two cases of the number of mass zones: $N=600$ and $N=1200$.
Both spectra were computed for the solution obtained on the time interval $t/\Pi_0\approx 1.3\times 10^3$.
An insignificant difference between these Fourier spectra allows us to conclude that long--term
cyclic variations of the pulsation amplitude are not the computational artefact because they
are almost independent of the number of mass zones of the hydrodynamic model.

\begin{figure}
 \centering
 \includegraphics[width=0.60\columnwidth,clip]{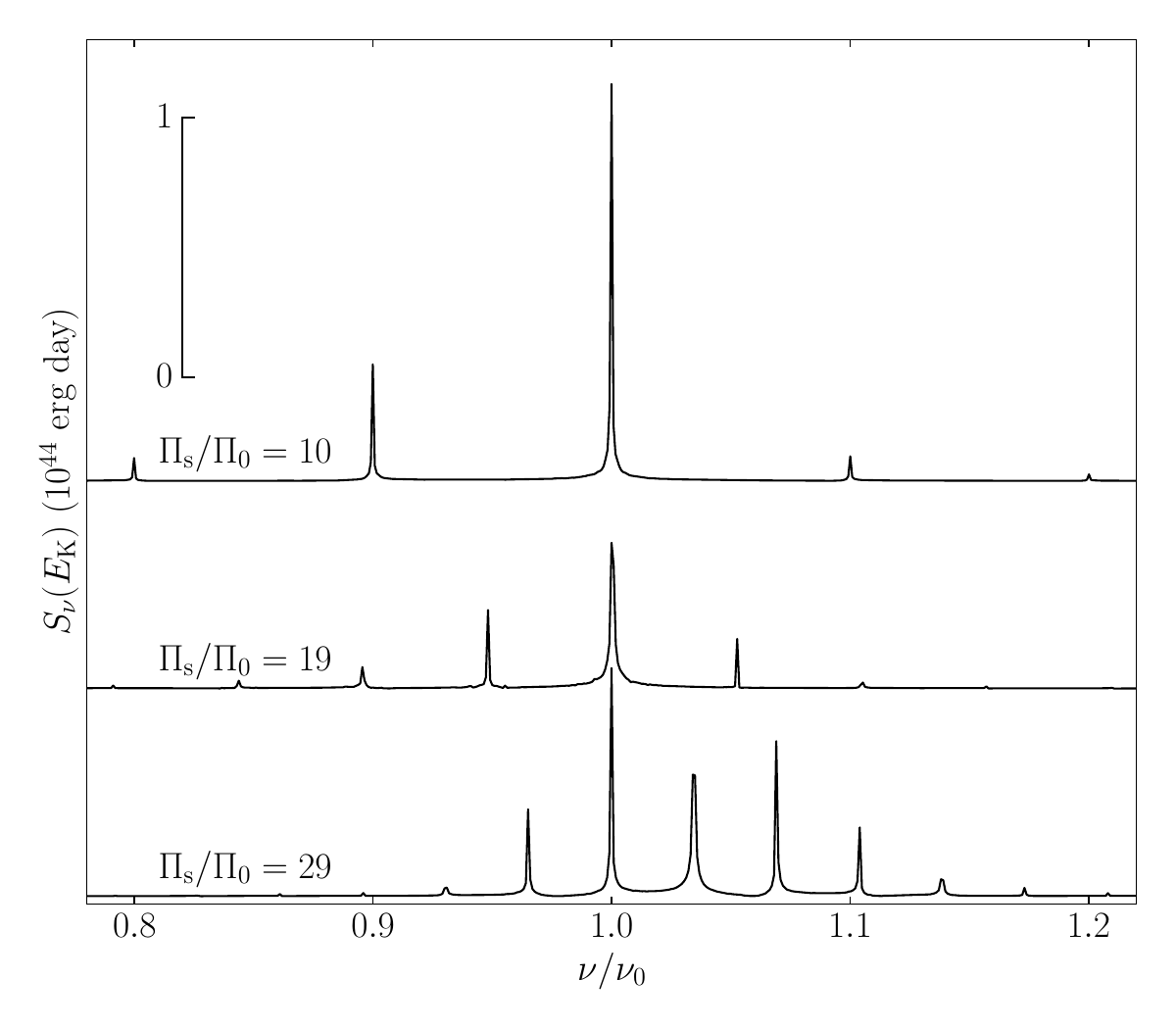}
 \caption{The kinetic enerhy power spectra $S_\nu(E_\mathrm{K})$ in the vicinity of the principal mode
          frequency $\nu_0 = 1/\Pi_0$ for three hydrodynamic models with ratios
          $\Pi_\mathrm{s}/\Pi_0 = 10$, 19 and 29.}
\label{fig8}
\end{figure}

An essential feature of the plots in Fig.~\ref{fig7} is that the principal maximum is split into
several equidistant frequency components, so that their superposition is responsible for
appearance of long--term cyclic variations of the pulsation amplitude.
The ratio $\Pi_\mathrm{s}/\Pi_0$ depends on the both the amplitude of peaks and frequency interval
between peaks.
This is clearly seen in Fig.~\ref{fig8} from power spectrum plots of three hydrodynamic models
with period ratios $\Pi_\mathrm{s}/\Pi_0 = 10$, 19 and 29.

Main characteristics of hydrodynamic models with secondary periodicity are listed in Table~\ref{tabl1}
in order of increasing frequency split $\Delta\nu/\nu_0$ which corresponds to decreasing period
ratio $\Pi_\mathrm{s}/\Pi$.
All models in Table~\ref{tabl1} are the fundamental mode pulsators.

\begin{table}
\caption{Main characteristics of hydrodynamic models with secondary periodicity.}
\label{tabl1}
\begin{center}

\begin{tabular}{ccccccc}
\hline
$M_\mathrm{ZAMS}/M_\odot$ & $M/M_\odot$ & $i_\mathrm{TP}$ & $\Pi, \textrm{day}$ & $\Pi_\mathrm{s}/\Pi$ & $\Delta\nu/\nu_0$ \\
\hline
3.0 &   2.490 &    8 &   469.2 &  30.8 &   0.0323 \\
3.0 &   2.423 &    9 &   461.5 &  28.9 &   0.0350 \\
3.0 &   2.476 &    8 &   472.5 &  28.3 &   0.0351 \\
3.0 &   2.461 &    8 &   474.4 &  27.0 &   0.0372 \\
3.0 &   2.452 &    8 &   472.4 &  27.0 &   0.0374 \\
3.0 &   2.420 &    9 &   467.4 &  26.8 &   0.0374 \\
3.0 &   2.417 &    9 &   471.9 &  25.7 &   0.0390 \\
2.8 &   2.181 &   12 &   431.6 &  22.8 &   0.0441 \\
2.6 &   1.969 &   14 &   401.9 &  19.2 &   0.0521 \\
3.0 &   2.010 &   11 &   385.1 &  15.7 &   0.0641 \\
2.8 &   2.055 &   13 &   405.4 &  14.3 &   0.0709 \\
3.0 &   2.234 &   10 &   476.6 &  10.0 &   0.1000 \\
\hline
\end{tabular}

\end{center}
\end{table}

Initially the secondary periodicity was found occasionally in several hydrodynamic models.
To clarify the connection between this phenomenon and the evolutionary changes of stellar structure
we calculated a few more hydrodynamic models of the evolutionary sequence $M_\mathrm{ZAMS}=3M_\odot$.
Thus, four models on the stage $i_\mathrm{TP}=8$ and three models on the stage $i_\mathrm{TP}=9$
represent two sequences of stellar models with continuous secondary periodicity.
The duration of the secondary periodicity stage is $t_\mathrm{sp}\approx 5.6\times 10^3$~yr for
$i_\mathrm{TP}=8$ and $t_\mathrm{sp}\approx 1.1\times 10^3$~yr for $i_\mathrm{TP}=9$.
The interflash interval is $\Delta t_\mathrm{TP}\approx 2.2\times 10^4$~yr so that the duration
of the second periodicity is as long as 25\% and 5\%, respectively.

\section*{conclusions}

Consistent calculations of stellar evolution and nonlinear stellar oscillations done in the present
work allow us to clarify some details about Mira variables.
Here we mention them.

In the early TP--AGB stage (i.e. during the first several thermal flashes) the red giant remains
stable against radial pulsations.
Evolutionary increase in the luminosity is accompanied by extension of the hydrogen and helium ionizing
zones where pulsations are excited.
Initially oscillations arise due to instability of the first overtone since the inner boundary of
ionization zones locates above the node of the first overtone with radius $r_\mathrm{n}/R\approx 0.77$,
where $R$ is the outer radius of the evolution model.
Models pulsating in higher order overtones were not found.
Enough dense grid of hydrodynamic models allowed us to determine the lower limit of the first
overtone period which increases with increasing stellar mass from $\Pi_{1,\min}\approx 80$~days
(the stellar mass $M=1.27M_\odot$, the evolutionary sequence $M_\mathrm{ZAMS}=1.5M_\odot$) to
$\Pi_{1,\min}\approx 120$~days ($M=2.60M_\odot$, $M_\mathrm{ZAMS}=2.8M_\odot$).

The existence of the upper limit of the first overtone period $\Pi_{1,\max}$ is due to the mode
switch to the fundamental mode.
The duration of mode switch in Mira variables comprises $\sim 10^2$ oscillation cycles (Fadeyev 2022)
and in comparison with the evolution timescale of TP--AGB stars can be treated as an instant process.
Moreover, in the vicinity of the mode switch the periods of the first overtone and fundamental mode
relate as $\Pi_1/\Pi_0 = 1/2$ so that together with the upper limit of the first overtone we obtain
the estimate of the lower limit of the fundamental mode period.
In the evolutionary sequences $1.5M_\odot\le M_\mathrm{ZAMS}\le 2.8M_\odot$ these limit values
increase with increasing stellar mass $M$ from  $\Pi_{1,\max} = 130$ and $\Pi_{0,\min} = 190$~days
for $M=0.96M_\odot$ to $\Pi_{1,\max} = 210$ and $\Pi_{0,\min} = 430$~days for $M=2.2M_\odot$.

Comparison of the results of the present work with those obtained earlier for Mira variable models
with $Z=0.014$ (Fadeyev 2023) allows us to conclude that the slope of the theoretical period--luminosity
relation increases with decreasing $Z$.
The metal dependence of the slope of empirical period--luminosity relations have been recently
demonstrated by Chibueze et al. (2020).

The theoretical period--luminosity relation was determined in the present study for bolometric
magnitudes and its slope (-4.33) is larger than the observational estimate in the K passband $\rho=-3.69$
(Whitelock et al., 2008).
One of the reasons of this difference might be the non--constant bolometric correction which
remains fairly uncertain.
On the other hand, the smaller slope of the theoretical dependence can probably be obtained for
$Z=0.008$ but this conclusion should be checked based of additional stellar evolution and
stellar pulsation computations.

Among nearly three hundred hydrodynamic Mira variable models calculated in the present study
twelve of them were found to have the kinetic energy power spectrum with split components in
the vicinity of the fundamental mode frequency.
In each model the frequency components are equidistant and for calculated hydrodynamic models
are in the range $0.03 \le \Delta\nu/\nu_0 \le 0.10$, where $\nu_0$ is the frequency of the
principal component.
The superposition of oscillations in the hydrodynamic models with the split kinetic energy
spectrum leads to the long--term cyclic variations of the pulsation amplitude.
The ratio of the secondary period $\Pi_\mathrm{s}$ to the fundamental mode period $\Pi_0$
depends of the frequency $\Delta\nu$ and ranges from $\Pi_\mathrm{s}/\Pi=10$ for $\Delta\nu/\nu_0 = 0.10$
to $\Pi_\mathrm{s}/\Pi=31$ for $\Delta\nu/\nu_0 = 0.03$.
The physical nature of the split of the kinetic energy spectrum remains, unfortunately, unclear.
To comprehend the origin of the pulsation spectrum split in radially pulsating stars
one should undertake a more detailed analysis of hydrodynamic models.
Perhaps this will allow us to understand the nature of the long secondary periodicity in
Mira variables.

\section*{references}

\begin{enumerate}
\item J.O.~Chibueze, R.~Urago, T.~Omodaka, Y.~Morikawa, M.Y.~Fujimoto, A.~Nakagawa, T.~Nagayama,
      T.~Nagayama, K.~Hirano, Publ. Astron. Soc. Japan \textbf{72}, 59 (2020).

\item A.A.~Cole, E.~Tolstoy, J.S.~Gallagher, T.A.~Smecker-Hane, Astron. J. \textbf{129}, 1465 (2005).

\item Yu.A.~Fadeyev, Astron. Lett. \textbf{39}, 306 (2013).

\item Yu.A.~Fadeyev, Mon. Not. R. Astron. Soc. \textbf{514}, 5996 (2022).

\item Yu.A.~Fadeyev, Astron. Lett. \textbf{49}, 722 (2023).

\item M.W.~Feast, Mon. Not. R. Astron. Soc. \textbf{211}, 51 (1984).

\item I.S.~Glass and T.~Lloyd~Evans, Nature \textbf{291}, 303 (1981).

\item I.S.~Glass and M.W.~Feast, Mon. Not. R. Astron. Soc. \textbf{199}, 245 (1982).

\item F. Herwig, Ann. Rev. Astron. Astrophys. \textbf{43}, 435 (2005).

\item C.D.~Huang, A.G.~Riess, S.L.~Hoffmann, C.~Klein, J.~Bloom, W.~Yuan, L.M.~Macri,
      D.O.~Jones, P.A.~Whitelock, S.~Casertano, and R.I.~Anderson, Astrophys. J. \textbf{857}, 67 (2018).

\item I.~Iben and A.~Renzini, Ann. Rev. Astron. Astrophys. \textbf{21}, 271 (1983).

\item P.~Iwanek, I.~Soszy\'nski, and S.~Koz\l{}owski, Astrophys. J. \textbf{919}, 99 (2021).

\item E.~Josselin, J.A.D.L.~Blommaert,  M.A.T.~Groenewegen, A.~Omont, and F.L.~Li,
      Astron. Astrophys. \textbf{357}, 225 (2000).

\item J.~Mould, A.~Saha, and S.~Hughes, Astrophys. J. Suppl. Ser. \textbf{154}, 623 (2004).

\item E.A.~Olivier and P.R.~Wood, Mon. Not. R. Astron. Soc. \textbf{362}, 1396 (2005).

\item B.~Paxton, R.~Smolec, J.~Schwab, A.~Gautschy, L.~Bildsten, M.~Cantiello, A.~Dotter,
      R.~Farmer, J.A.~Goldberg, A.S.~Jermyn, S.M.~Kanbur, P.~Marchant, A.~Thoul, R.H.D.~Townsend, W.M.~Wolf,
      M.~Zhang, and F.X.~Timmes, Astrophys. J. Suppl. Ser. \textbf{243}, 10 (2019).

\item W.R.J.~Rolleston, C.~Trundle, and P.L.~Dufton, Astron. Astrophys. \textbf{396}, 53 (2002).

\item M.~Trabucchi, P.R.~Wood., J.~Montalb\'an, P.~Marigo, G.~Pastorelli, L.~Girardi,
      Mon. Not. R. Astron. Soc. \textbf{482}, 929 (2019).

\item M.~Trabucchi, P.R.~Wood, N.~Mowlavi, G.~Pastorelli, P.~Marigo, L.~Girardi and T.~Lebzelter),
      Mon. Not. R. Astron. Soc. \textbf{500}, 1575 (2021).

\item Y.~Tuchman, N.~Sack, and Z.~Barkat, Astrophys. J. \textbf{219}, 183 (1978).

\item P.A.~Whitelock, F.~Marang, and M.W.~Feast, Mon. Not. R. Astron. Soc. \textbf{319}, 728 (2000).

\item P.A.~Whitelock, M.W.~Feast and F.~Van~Leeuwen, Mon. Not. R. Astron. Soc. \textbf{386}, 313 (2008).

\item P.A.~Whitelock, J.W.~Menzies, M.W.~Feast F.~Nsengiyumva, and N.~Mat\-su\-naga,
      Mon. Not. R. Astron. Soc. \textbf{428}, 2216 (2013).

\item L.A.~Willson, Ann. Rev. Astron. Astrophys. \textbf{38}, 573 (2000).

\item P.R.~Wood, Astrophys. J. \textbf{190}, 609 (1974).

\item P.R.~Wood and D.M.~Zarro, Astrophys. J. \textbf{247}, 247 (1981).

\item P.R.~Wood, C.~Alcock, R.A.~Allsman, D.~Alves, T.S.~Axelrod, A.C.~Becker, D.P.~Bennett, K.H.~Cook,
      A.J.~Drake, K.C.~Freeman, K.~Griest, L.J.~King, M.J.~Lehner, S.L.~Marshall, D.~Minniti, B.A.~Peterson,
      M.R.~Pratt, P.J.~Quinn, C.W.~Stubbs, W.~Sutherland, A.~Tomaney, T.~Vandehei, D.L.~Welch,
      \textit{Asymptotic Giant Branch Stars},
      Ed. by T. Le Bertre, A~Lebre, and C.~Waelkens,
      (IAU Symposium 191, 1999), p. 151.

\item P.R.~Wood, Publ. Astron. Soc. Australia \textbf{17}, 18 (2000).

\item W.~Yuan, L.M.~Macri, A.~Javadi, Z.~Lin, and J.Z.~Huang, Astron. J. \textbf{156}, 112 (2018).

\end{enumerate}

\end{document}